\documentclass[10pt,twoside]{article}
\usepackage[english]{babel}
\fontencoding{T1}
\usepackage{amsmath}
\usepackage{amssymb}

\usepackage{diagxy}
 
\title{Kolmogorov complexity and the geometry of Brownian motion}

\author{Willem L. Fouch\'{e}\\
\it Department of Decision Sciences,\\\it School of Economic Sciences\\\it University of South
Africa, PO Box 392, 0003 Pretoria, South
Africa\\fouchwl@gmail.com}
\date{}
\newtheorem{theorem}{Theorem}
\newtheorem{proposition}{Proposition}
\newtheorem{lemma}{Lemma}
\newtheorem{corollary}{Corollary}
\newtheorem{definition}{Definition}

\begin{document}

%\thanks {The research is based upon work supported by the National Research Foundation (NRF) of South Africa. Any opinion, findings and conclusions or recommendations expressed in this material are those of the author and therefore the NRF does not accept any liability in regard thereto.}  \date{}
\maketitle
%\begin{center}
%\vspace*{1in}
%{\it To the memory of my parents Hennie en Charlotte}%\end{center} 
%\end{center}
\begin{abstract}
 In this paper, we continue the study of the  geometry of Brownian motions which are encoded by Kolmogorov-Chaitin random reals (complex oscillations). We unfold Kolmogorov-Chaitin complexity in the context of Brownian motion and specifically to  phenomena emerging from the random geometric patterns generated by a Brownian motion.

\end{abstract}

\section{Introduction}
\label{Introduction}
\begin{quote}
Finally, we would like to comment on the hidden role of Kolmogorov complexity in the real life of classical computing ....

The inherent tension, incompatability of shortest descriptions with most-economical  algorithmical processing, is the central issue of any computability theory.

The place-value notation of numbers that played such a great role in the development of human civilizations is the ultimate  system of short descriptions that bridges the abyss.  Kolmogorov complexity goes far beyond this point. (\cite{manin:1} p 327.)

\end {quote} 
It is well-known that the notion of randomness, suitably refined, goes a a long way in dealing with this tension. (See, for example, \cite{ch:2,marlo:1,nie:1}.)
In this paper, we continue to explore this interplay between short descriptions and randomness in the context of Brownian motion and its associated geometry. In this way one sees how random phenomena associated with the geometry of Brownian motion, are implicitly enfolded in each real number which is
complex in the sense of Kolmogorov. These random phenomena range from fractal geometry, Fourier analysis and non-classical noises in quantum physics.

%A function on the unit interval is  {\it
%algorithmically random} if it fails all effective (Martin-L\" of) statistical tests, now expressed in terms of
%the statistical events associated with Brownian motion on the unit
%interval.  The class of functions corresponds exactly, in  the language of Weihrauch \cite{wei:1,wei:2},  G\'acs \cite{gac:1} and Hoyrup-Rojas \cite{horo:1},  in the context of algorithmic randomness, to the Martin-L\" of random elements of the computable measure space 
% \[ \mathcal{R}= (C_0[0,1],d,B,W),\] where $C_0[0,1]$ is the set of the continuous functions on the unit interval that vanish at the origin, $d$ is the metric induced by the supremum norm, $B$ is the countable set of piecewise linear functions vanishing at the origin with slopes and  points of non-differentiability all rational numbers and where $W$ is the Wiener measure (which is a uniformly computable measure relative to a recursive enumeration of the elements of $B$).

We study in this paper algorithmically random Brownian motion, the representations of which were
also called  {\it complex
oscillations} in \cite{fo:1,fo:2} for example. This terminology was suggested to the author by the following Kolmolgorov theoretic interpretation  of this notion by Asarin and Prokovskii
\cite{ap:1}, who are the pioneers of this theme. One can characterise a Brownian motion which is algorithmically random (or, equivalently Martin-L\" of random)
 as an effective and uniform limit
of a sequence $(x_n)$ of ``finite random walks'', where, moreover,
each $x_n$ can be encoded by a finite binary string $s_n$ of length
$n$, such that the (prefix-free) Kolmogorov complexity, $K(s_n)$, of $s_n$
satisfies, for some constant $d > 0$, the inequality $K(s_n) > n-d$
for all values of $n$. 

Two other characterisations of the class of complex oscillations were developed by the author in \cite{fo:1} and \cite{fo:2}. In \cite{fo:1} the class of complex oscillations were described in terms of effective subalgebras of the Borel $\sigma$-algebra on $C[0,1])$ (with the supremum norm topology). The central idea here was to effectivise the Donsker invariance principle, i.e., to focus on Brownian motion as a scaling limit of finitary random walks. In \cite{fo:2} they are shown to be exactly those real-valued continuous functions on the unit interval that can be computed from Martin-L\" of-random reals (relative to the Lebesgue measure) by means of an associated Franklin-Wiener series.  Here, the guiding motif was the fact that one can, as is well-known, also think of Brownian motion as a linear superposition of deterministic oscillations with normally distributed random amplitudes.  Further applications and developments  of this idea can be found in the papers \cite{fo:4,fo:5,horo:1, paulpot:1,KHNe:2}. A sharper version, from a computational point of view,  of the main result in \cite{fo:2}  has recently been developed by George Davie and the author \cite{dafo:1}. 

 Countable dense random sets  arise naturally in the theory of Brownian motion \cite{tsi:1}, in non-classical noises \cite{tsi:3} and the understanding of percolation phenomena in statistical physics (see \cite {tsi:3}, \cite{camfo:1},  for example).  It is an interesting fact that  the study of countable dense random sets quite naturally brings one in contact with studying random processes over spaces which are not even Polish. One has to do probability theory over orbit spaces under the action of the group $S_\infty$, which is the symmetry group of a countable set, on the space of all injections of $\mathbb{N}$ into the unit interval. These are examples of what Kechris \cite{kechris:1} referred to as singular spaces of Borel cardinality $F_2$. 

  In \cite{tsi:1} Tsirelson develops a very powerful approach to random processes over these singular spaces and his results imply that  the Kechris-singularity  manifests in very concrete and interesting statistical properties of countable dense random sets and new aspects of Brownian motion.

Tsirelson \cite{tsi:1}  shows that the  minimizers of a Browian motion are, in the language of \cite{tsi:3}, instances of so-called  stationary local random dense countable sets over the white noise and that they play a pivotal role in the understanding of non-classical noises. 

 This work suggested to the author the problem of  constructing  the minimizers of Brownian motion directly from an unbiased coin-tossing experiment.  This can be seen as an extension of \cite{fo:2} where a  generic Brownian motion was constructed from a generic point in the unit interval. We shall again adopt the viewpoint of Kolmogorov complexity to define what we mean by the word generic. In this way, we shall be able to find $\Sigma_3^0$ definitions, within the arithmetical hierarchy,  for countable dense random sets, which can be considered to be ``generic" countable dense sets of reals and moreover symmetrically random over white noise. We provide an explicit  computable enumeration of the elements of such sets relative to Kolmogorov-Chaitin-Martin-L\" of  random real numbers. This opens the way to relate certain non-classical noises to Kolmogorov complexity. For example, the work of the present paper enables one to represent Warren's splitting noise (see \cite{tsi:3}) directly in terms of infinite binary strings which are Kolmogorov-Chaitin-Martin-L\" of random. This line of thought will also be pursued in a sequel to this paper.

In this sequel to this paper, we shall  study the  images of certain $\Pi_2^0$  perfect sets of Hausdorff dimension zero under a complex oscillation. We have given a sketch  in  the extended abstract \cite{fo:5} of a proof that there are instances of such sets  where such images under complex oscillations have  elements all of which are linearly independent over the field of rational numbers. In Fourier analysis, these sets are called sets of independence. (See, for example, Chapter 5  of Rudin's book \cite{rud:1}, pp 97-130.) We shall provide a generalisation of this result and show in fact that one can obtain sets via complex oscillations which are linearly independent over the field of recursive real numbers.  Moreover, all the elements in these images are non-computable. %We discuss the definability of these sets within the recursion-theoretic hierarchy, by exploiting the recursive isomorphism constructed in \cite{fo:2} between the Kolmogorov-Chaitin-Martin-L\" of random reals and the class of suitably encoded versions of  complex oscillations. 

%By these means the author wishes to depict  the incredibly  rich geometry that is enfolded in every Kolmogorov-Chaitin  random binary string $\alpha$ by merely regarding such an $\alpha$ as an encoding of a complex oscillation or, equivalently, of an (effectively) generic Brownian motion.  More instances of this phenomenon were developed by the  author \cite{fo:2,fo:4,fo:5} and by  Kjos-Hanssen and Nerode in \cite{KHNe:1, KHNe:2}.
% The arguments in this paper rely heavily on \cite{fo:1}, \cite{fo:2},  as well as on the beautiful constructions of Kahane in \cite{kah:1}.

The author is grateful to the Department of Mathematics at the Corvinus University, Budapest, for hosting my frequent visits to the department and for sharing with me so much of the subtleties of  measure theory and stochastic processes.

The  research in this paper has been supported by the National Research Foundation (NRF) of South Africa and  by the European Union  grant agreement PIRSES-GA-2011-2011-294962 in Computable Analysis (COMPUTAL).

Many thanks are due to the referee whose remarks led to a significant strengtening of Theorem \ref{theorem:description}.

\section{Preliminaries and statements of the main theorems}
%\usepackage{amsmath}
%\usepackage{amsfonts}
%\usepackage{amssymb}
%$\boxplus_mK$$\biguplus_mK$ 
%\[\bfig
%\Vtriangle/>`>`>/[A`B`C;a`b`c]
%\efig\]

A Brownian motion on the unit interval is a real-valued function $(\omega,t) \mapsto X_\omega(t)$ on $\Omega \times [0,1]$, where $\Omega$ is the underlying space of some probability space, such that $X_\omega(0)=0$ a.s. and for $t_1 < \ldots < t_n$ in the unit interval, the random variables $X_\omega(t_1),X_\omega(t_2)-X_\omega(t_1), \cdots, X_\omega(t_n)-X_\omega(t_{n-1})$ are statistically independent and normally distributed with means all $0$ and variances $t_1,t_2-t_1,\cdots,t_n-t_{n-1}$, respectively. We say in this case that the Brownian motion is parametrised by $\Omega$. Alternatively, the map $X$ defines a Brownian motion iff for $t_1 < \ldots <t_n$ in the unit interval, the random vector $(X_\omega(t_1), \cdots, X_\omega(t_n))$ is Gaussian with correlation matrix $(\mbox{min}(t_i,t_j):1 \leq i,j \leq n)$.

It is a fundamental fact that any Brownian motion has a ``continuous version''. This means the following: Write $\Sigma$ for the $\sigma$-algebra of Borel sets of $C[0,1]$ where the latter is topologised by the uniform norm topology. There is a probability measure $W$ on $\Sigma$ such that for $0\leq t_1 < \ldots <t_n \leq 1 $ and for a Borel subset $B$ of $\mathbf{R}^n$, we have
\[ P(\{\omega \in \Omega:(X_\omega(t_1), \cdots, X_\omega(t_n)) \in B \})
=W(A),\]
where
\[A=\{x \in C[0,1]: (x(t_1), \cdots, x(t_n)) \in B\}.\]
 The measure $W$ is known as the {\it Wiener measure}. We shall usually write $X(t)$ instead of $X_\omega(t)$.

In the sequel, we shall denote by $(0,1)^\infty$ the Borel space consisting of the product of countably many copies of the unit interval and with Borel structure being given by the natural product structure which is induced by the standard Borel structure on the unit interval. We write $(0,1)^\infty_{\neq}$ for the Borel subspace consisting of the infinite sequences in the unit interval which are pairwise distinct.

We write $S_\infty$ for the symmetric group of a countable set (which we can take to be $\mathbb{N}$). We place on $S_\infty$ the pointwise topology. We thus give $S_\infty$ the subspace topology under the embedding of $S_\infty$ into the Baire space $\mathbb{N}^\mathbb{N}$. The group $S_\infty$ acts naturally (and continuously) on $(0,1)^\infty_{\neq}$ as follows:
$$\sigma.(u_j: j \geq 1) := (u_{\sigma^{-1}(j)}: j \geq 1),$$
for all $(u_j) \in (0,1)^\infty_{\neq}$ and $\sigma \in S_\infty$ (the logical action). The orbit space under this action is denoted by $(0,1)^\infty_{\neq}/S_\infty$. We place a Borel structure on this space via the topology induced by the canonical mapping 
$$\pi:(0,1)^\infty_{\neq} \longrightarrow (0,1)^\infty_{\neq}/S_\infty.$$

Let $\Omega$ be standard Borel space. A {\it strongly } random countable set in the unit interval is a measurable mapping  $X: \Omega \rightarrow (0,1)^\infty_{\neq}/S_\infty$ that factors through some (traditional) random sequence $Y$ as shown:
\[\bfig
\Vtriangle/>`>`>/[\Omega`(0,1)^\infty_{\neq}.`(0,1)^\infty_{\neq}/S_\infty;Y`X`\pi]
\efig\]
One can think of $X$ as a random countable {\it set} induced via $S_\infty$-equivalence, by a random {\it sequence} $Y$, both in the unit interval. In the sequel, we shall sometimes denote the Borel space $(0,1)^\infty_{\neq}/S_\infty$ by $CS(0,1)$.

As noted by Tsirelson \cite{tsi:1} the natural question as to whether any measurable $X: \Omega \rightarrow (0,1)^\infty_{\neq}/S_\infty$ factors through some $Y$ as above, is an open problem. % My guess is no,  but this is nothing but a guess. 

The following fundamental theorem of Tsirelson's explains exactly what it means for two strongly random countable sets to be ``statistically similar".

\begin{theorem} (\cite{tsi:1}). For standard measure spaces $(\Omega_1,P_1)$ and $(\Omega_2,P_2)$, let, for $i=1,2$,  there be,   some $P_i$-measurable strongly random set $X_i:\Omega _i\rightarrow CS(0,1)$  such that the induced probability distributions on $CS(0,1)$ are the same, i.e., for every Borel subset $\Sigma$ of $CS(0,1)$ it is the case that
$$P_1(X_1^{-1}(\Sigma))=P_2(X_2^{-1}(\Sigma)).$$ Then there is a probability distribution $\mathbb{P}$ on $\Omega_1 \times \Omega_2$ such that the marginal of $\mathbb{P}$ to $\Omega_i$ is $P_i$, and moreover,  for $\mathbb{P}$ almost all $(\omega_1,\omega_2) \in \Omega_1 \times \Omega_2$  it is the case that $$X_1(\omega_1)=X_2(\omega_2).$$ 
\end{theorem}

The statement ``the marginal of $\mathbb{P}$ to $\Omega_i$ is $P_i$", means that for measurable $\Sigma_i \subset \Omega_i\;\; i=1,2$:

$$\mathbb{P}(\Sigma_1 \times \Omega_2)=P_1(\Sigma_1);$$
and
$$\mathbb{P}(\Omega_1 \times \Sigma_2)=P_2(\Sigma_2).$$

We say in this case that the strongly random sets $X_1$ and $X_2$ are {\it statistically similar} relative to the probabilities $P_1,P_2$ and we simply  write  $X_1 \sim X_2$.

 A strongly random countable set $X:\Omega \rightarrow CS(0,1)$ is said to be {\it generic}  relative to a probability measure $P$ on $\Omega$ if the following is true: \begin{quote} If  $B$ is a Borel subset of the unit interval  such that $\lambda(B) > 0$, then $P$- almost surely, $B \cap X\neq \emptyset$. On the other hand, if $\lambda(B) = 0$, then $P$-almost surely, $B \cap X = \emptyset$. Equivalently, if $C$ is a Borel set such that $\lambda(C)=1$, then, almost surely, $X \subset C$. \end{quote}
Here we have written $\lambda$ for the Lebesgue measure on the unit interval. Note that if $X$ is generic and if $Y \sim X$, then $Y$ too is generic.
%Independence condition.

A partial converse of this statement can be found in \cite{tsi:1}: If  $X_1,X_2$ are both strongly random, each satisfying what Tsirelson calls the ``independence condition" relative to a probability measure $P_i$, and each being almost surely dense in the unit interval, then  they are statistically similar provided they are both generic!!  (Tsirelson 2006).

Write $\lambda^\infty$ for the product measure on $(0,1)^\infty$ which is the countable product of the Lebesgue measure $\lambda$ on the unit interval and write $\Lambda$ for the measure on $CS(0,1)$ 
which is the pushout of $\lambda^\infty$ under $\pi$. In other words, for a Borel subset $\Sigma$ of $CS(0,1)$,
$$\Lambda(\Sigma) =\lambda^\infty(\pi^{-1}\Sigma).$$

Write $U:(0,1)^\infty \rightarrow CS(0,1)$ for the strongly random set as defined by the following commutative diagram:
%\rightarrow (0,1)^\infty/S_\infty$ that factors via some (traditional) random sequence $Y$ as shown:
\[\bfig
\Vtriangle/>`>`>/[(0,1)^\infty`(0,1)^\infty.`CS(0,1)=(0,1)^\infty_{\neq}/S_\infty;Id`U`\pi]
\efig\]

Then $U$ is almost surely dense and generic. In statistics $U$ is a model of an unordered uniform infinite sample. Moreover, it follows from the Hewitt-Savage theorem, that for every Borel subset $\Sigma$ of $CS(0,1)$, it is the case that 
\begin{equation}
\Lambda(\Sigma) \in \{0,1\}.
\label{equation:Hewitt-Savage}
\end{equation}
Note that $\Lambda$ is non-atomic. Consequently,   $CS(0,1)$ is not a Polish space! We shall refer to the strongly random set $U$ as the {\it uniform random set}.

If $X$ is a continuous function on the unit interval, then a {\it local minimizer} of $X$ is a point $t$ such that there is some closed interval $I \subset[0,1]$ containing $t$ such that the function $X$ assumes its minimum value on $I$ at the point $t$. We denote by $\verb=MIN=(X)$ the set of local minimizers of $X$. It is well-known that if $X$ is a continuous version of Brownian motion on the unit interval, then $\verb=MIN=(X)$ is almost surely a dense and countable set and that all the local minimizers of $X$ are {\it strict}. This means that, for each closed subinterval $I$ of the closed unit interval, there is a unique $\nu \in I$ where the minimum of $X$ on $I$ is assumed.  This, as will be explained in this paper, has the implication that there is a subset $\Omega_0$ of $C[0,1]$ of full Wiener measure such that one can define a measurable mapping  $\verb=min=: C[0,1] \supset \Omega_0  \longrightarrow (0,1)^\infty_{\neq} $ in such a way that the composition of $\verb=min=$ with the projection $\pi$ will define a measurable mapping $X \mapsto \verb=MIN=(X)$. In the sequel this strongly random set will be denoted by $\verb=MIN=$. To summarise, we have the following commutative diagram:%Make explicit that it is similar to uniform random set $U$. Recall Hewitt-Savage and deduce $0-1$-law for the minimizers. 
%\rightarrow (0,1)^\infty/S_\infty$ that factors via some (traditional) random sequence $Y$ as shown:

\[\bfig
\Vtriangle/>`>`>/[C\hbox{[}0,1\hbox{]} \supset \Omega_0`(0,1)^\infty_{\neq}.`(0,1)^\infty_{\neq}/S_\infty;\min`\hbox{MIN}`\pi]
\efig\]

The next theorem of Tsirelson (2006) says essentially that the local minimizers of a Brownian motion is a generic countable dense random set. (It is quite trivial to show that it satisfies the independence property.)
\begin{theorem}\cite{tsi:1}.
 If $X$ is a  continuous version of Brownian motion on the unit interval and $B$ is a Borel subset of the unit interval such that $\lambda(B) > 0$, then almost surely, $B \cap \verb=MIN=(X) \neq \emptyset$. On the other hand, if $\lambda(B) = 0$, then almost surely, $B \cap \verb=MIN=(X) = \emptyset$. In particular, if $\lambda(C)=1$, then, almost surely, $\verb=MIN=(X) \subset C$.
\label{theorem:tsirelson}
\end{theorem}
It follows that  any generic countable dense random set with the independence property will be statistically similar to the random set of minimizers of a Brownian motion. In particular
\begin{equation}
\verb=MIN= \sim U.
\label{equation:similar}
\end{equation}
 %Deduce analogue of Hewitt-Savage for minimizers. Justify tha analysis of the more complicated $N$ in stead of $U$. Crucially important. Else what follows is unnecessarily complicated. Intuitive idea: $N$ intersects each interval in a canonical point enabling new processses to be defined over the random set. I think this is Tsirelson´s intuition. Link with Warren´s idea of stickiness, splitting, associated with   bernoulli processes over the minimizers, non-classical notions of independence. 

The set of words over the alphabet $\{0,1\}$ is denoted by $\{0,1\}^*$. If $a \in \{0,1\}^*$, we write $|a|$ for the length of $a$. If $\alpha=\alpha_0\alpha_1\ldots$ is an infinite word over the alphabet $\{0,1\}$ , we write $\overline{\alpha}(n)$ for the word $\prod_{j<n}\alpha_j$. We use the usual recursion-theoretic terminology $\Sigma_r^0$ and $\Pi_r^0$ for the arithmetical subsets of \(\mathbb{N}^k \times \{0,1\}^{\mathbb{N}\times l},\;k,l \geq 0 \). (See, for example, \cite{hin:1}). We again write $\lambda$ for the Lebesgue probability measure on $\{0,1\}$. For a binary word $s$ of length $n$, say, we write $[s]$ for the ``interval'' $\{\alpha \in \{0,1\}^\mathbb{N}: \overline{\alpha}(n) = s \}$. A sequence $(a_n)$ of real numbers converges {\it effectively} to $0$ as $n \rightarrow \infty$ if for some total recursive $f:\mathbb{N} \rightarrow \mathbb{N}$, it is the case that $|a_n| \leq (m+1)^{-1}$ whenever $n \geq f(m)$.

For any finite binary word $a$ we denote its (prefix-free) Kolmogorov complexity by $K(a)$.  Recall that an infinite binary string $\alpha$ is Kolmogorov-Chaitin complex if 

\begin{equation}
\exists_d \forall_n\;K(\overline{\alpha}(n)) \geq n-d.
\label{eq:Kolmogorov}
\end{equation}
In the sequel, we shall denote this set by $KC$ and refer to its elements as $KC$-strings. (See, e.g., \cite{ch:2}, \cite{marlo:1} or \cite{nie:1} for more background.)

%We next survey the results from \cite{ap:1}, \cite{fo:1} and \cite{fo:2} which will play an important role in this discussion.  

For $n \geq 1$, we write $C_n$ for the class of continuous functions on the unit interval that vanish at $0$ and are linear with slopes $\pm \sqrt{n}$ on the intervals \([(i-1)/n,i/n]\;,i=1,\ldots ,n\). With every $x \in C_n$, one can associate a binary string $a =a_1\cdots a_n$ by setting $a_i=1$ or $a_i =0$ according to whether $x$ increases or decreases on the interval $[(i-1)/n,i/n]$. We call the sequence $a$ the code of $x$ and denote it by $c(x)$. The following notion was introduced by Asarin and Prokovskii in \cite{ap:1}.
\begin{definition}

A sequence $(x_n)$ in $C[0,1]$ is \emph{ complex}  if $x_n \in C_n$ for each $n$ and there is a constant $d > 0$ such that $K(c(x_n)) \geq n-d$ for all $n$. A function $x \in C[0,1]$ is a \emph{complex oscillation} if there is a complex sequence $(x_n)$ such that $\|x-x_n\|$ converges effectively to $0$ as $n \rightarrow \infty$.\
\label{definition:ap}
\end{definition}

The class of complex oscillations is denoted by $\mathcal{C}$. It was shown by Asarin and Prokovskii \cite{ap:1} that the class $\mathcal{C}$ has Wiener measure $1$.  In fact, they implicitly  showed that the class corresponds exactly, in the broad context  and modern language of Hoyrup and Rojas \cite{horo:1},  to the Martin-L\" of random elements of the computable measure space  (\cite{wei:1, wei:2, gac:1})
\begin{equation}
 \mathcal{R}= (C_0[0,1],d,B,W),
\label{eq:AP}
\end{equation}
where $C_0[0,1]$ is the set of  continuous functions on the unit interval that vanish at the origin, $d$ is the metric induced by the uniform norm, $B$ is the countable set of piecewise linear functions $f$ vanishing at the origin with slopes and  points of non-differentiability all rational numbers and where $W$ is the Wiener measure. 

  For recent refinements of this result, the reader is referred to the work of  Kjos-Hanssen and Szabados \cite{KHSza:1}.  They note  that Brownian motion and scaled,  interpolated simple random walks can be
jointly embedded in a probability space in such a way that almost surely, the $n$-step walk is, with respect to the uniform norm, 
within a distance $ O(n^{-\frac{1}{2} } \log n)$ of the Brownian path, for all but finitely many 
positive integers $ n$.  In the same paper, Kjos-Hanssen and Szabados show that, almost surely, their constructed  sequence $(x_n)$ of $n$-step walks is complex  in the sense of  Definition  \ref{definition:ap} and all Martin-L\" of random paths   have such
an incompressible close approximant. This strengthens a result of Asarin \cite{as:1}, who obtained
instead the bound $O(n^{-\frac{1}{6}} \log n)$. 

 The following theorem  can be extracted from \cite{fo:2}:

\begin{theorem} There is a bijection $\Phi:KC \rightarrow\mathcal{C}$ and a  uniform algorithm that, relative to any $KC$-string $\alpha$, with input a dyadic rational number $t$ in the unit interval and a natural number $n$, will output the first $n$ bits of the the value of the complex oscillation $\Phi(\alpha)$ at  $t$. 
 \label{theorem:isomorphism}
\end{theorem}

The construction in \cite{fo:2} of the complex oscillation $\Phi(\alpha)$ from a given $\alpha \in KC$ is as follows. Beginning with $\alpha \in KC$ we can construct a sequence of reals $\xi_0,\,\xi_1,\,\xi_{jn},\,j\geq 1,\,0\leq n<2^j$; the sequence is computable in $\alpha, j$ and $n$.  Thereafter, we recursively find $x(n/2^j)$ for $n,j \in \mathbb{N}$ with $n \leq 2^j$ from the $\xi$-sequence by solving the equations
$$x(1)=\xi_0,\hspace{0.5cm}2x(\frac{1}{2})=\xi_0+\xi_1,$$
and
$$2x\left(\frac{2n+1}{2^{j+1}}\right)=2^{-j/2}\xi_{jn}+x\left(\frac{n+1}{2^j}\right)+x\left(\frac{n}{2^j}\right).$$
By the arguments in \cite{fo:2}, the complex oscillation $\Phi(\alpha)$ associated with a given $\alpha \in KC$ turns out to be the unique continuous function which assumes, for every dyadic rational $d$,  the value $x(d)$.
In this way, one can effectively compute any finite initial segment of the value of $\Phi(\alpha)$ at a given dyadic rational number from some initial segment of $\alpha$.  It also follows from the construction in \cite{fo:2} that
\begin{equation}
\Phi(\alpha)=-\Phi(\hat{\alpha}).
\label{equation:fundamental symmetry}
\end{equation}
Here $\hat{\alpha}$ denotes the binary string obtained from $\alpha$ by replacing each bit $\alpha_i$ of $\alpha$ by $1-\alpha_i$.

The mapping $\Phi$ is also measure-preserving in the following sense:  Let $B$ be a Borel subset of $C[0,1]$. Then
$$\lambda(\alpha \in KC: \Phi(\alpha) \in B)=W(B).$$

%Consider any effective enumeration $t_0,t_1,\ldots,$ without repetition, of the dyadic rationals in the unit interval.  To every $x\,\,\epsilon\,\,C[0,1]$ one can associated the $\omega\times\omega$ array having the dyadic expansion of $x(t_i)$ as its $i$th row.  By using any recursive bijection between $\omega^2$ and $\omega$, one can represent the array associated to the continuous function $x$ as a single binary string, $E(x),$ say.  Set $\mathcal{E}=\left\{E(x):\,x\,\epsilon\,\mathcal{C}\right\}.$  In \cite{fo:2} it is shown that the map $\Phi$ induces a partial recursive $\phi:\left\{0,1\right\}^{\omega}\rightarrow\left\{0,1\right\}^{\omega}$, the restriction of $\psi$ to $\mathcal{E}$ is the inverse to $\phi.$  In this sense the sets KC and $\mathcal{E}$ are recursively isomorphic. 

%Write $M$ for the bijection that associates every encoded version of a complex oscillation the unique continuous function of which it is the code, we thus find the following commutative diagram.

%\[\bfig
%\Vtriangle/>`>`<-/[KC`\mathcal{C}.`\mathcal{E};\Phi`\phi`\nu]
%\efig\]

Let $\mathcal{N}$ be the function that associates with every $x \in \mathcal{C}$, the set of local minimizers of $x$. We shall discuss the measurability and computability of $\mathcal{N}$ in Section \ref{section:local minimizers} of this paper. Thus $\mathcal{N}$  is the restriction of $\verb=MIN=$ to $\mathcal{C}$.
We then define the function $$\mathcal{MIN}:KC \longrightarrow (0,1)^\infty_{\neq}/S_\infty $$ by
$$\alpha \mapsto \verb=MIN=(\Phi(\alpha));$$
this means that the diagram
\[\bfig
\Vtriangle/>`>`<-/[KC`{(0,1)^\infty_{\neq}/S_\infty}`\mathcal{C};\mathcal{MIN}`\Phi`\mathcal{N}]
\efig\]
 commutes.
It follows from (\ref{equation:Hewitt-Savage}), (\ref{equation:similar}) and the fact that $\Phi$ is measure-preserving, that, for every Borel subset $\Sigma$ of $(0,1)^\infty_{\neq}/S_\infty$:
$$\lambda(\alpha \in KC: \mathcal{MIN}(\alpha)\in \Sigma) \in \{0,1\}.$$
(The zero-one law for the minimizers of complex oscillations.)

What is essentially at stake here is the Hewitt-Savage theorem together with the statistical similarity of three strongly random sets:
$$U \sim \mathcal{N} \sim \mathcal{MIN}.$$
%\begin{theorem}  (F 2000, F, Davie 2011.) There is a bijection $\Phi:KC \rightarrow\mathcal{C}$ and a  uniform algorithm that, relative to any $KC$-string $\alpha$, with input a rational number $t$ in the unit interval and a natural number $n$, will output the first $n$ bits of the the value of the complex oscillation $\Phi(\alpha)$ at the value $t$.  Moreover, the bijection $\Phi$ is totally effective and explicit relative to an additional oracle with access to an uNpper bound of the compressibilty coefficient of the $KC$-string $\alpha$. 
{\bf Remark.} It would be interesting to better understand the Borel subsets $\Sigma$ of $(0,1)^\infty_{\neq}/S_\infty$ having $\Lambda$ measure one such that $\mathcal{MIN}(\alpha) \in \Sigma$ for all $\alpha \in KC$
%\end{theorem}
In this paper we shall prove
\begin{theorem}
 There is a uniform procedure that, relative to a given $\alpha \in KC$, will yield, for any closed dyadic subinterval $I$ of the unit interval, a sequence $t_1,t_2, \ldots$ of rationals in $I$ that converges to the (unique) local minimizer of the complex oscillation, $\Phi(\alpha)$, in $I$. Moreover all the local minimizers of a complex oscillation are non-computable real numbers.
\label{theorem:algorithm}
\end{theorem}
We shall also prove
\begin{theorem}
There  is a $\Sigma_3^0$ predicate $C(\alpha,\nu)$ over $\{0,1\}^{\mathbb{N}}\times \{0,1\}^\mathbb{N}$ such that for $\alpha, \nu \in \{0,1\}^\mathbb{N}$
\[ C(\alpha,\nu) \Longleftrightarrow \nu \in \verb=MIN=(\Phi(\alpha)) \wedge \alpha \in KC .\]
\label{theorem:description}
\end{theorem}
%In this paper, we shall  study the  images of certain $\Pi_2^0$  perfect sets of Hausdorff dimension zero under a complex oscillation. We have given a sketch  in  the extended abstract \cite{fo:5} of a proof that there are instances of such sets  where such images under complex oscillations have  elements all of which are linearly independent over the field of rational numbers. In Fourier analysis, these sets are called sets of independence. (See, for example, Chapter 5  of Rudin's book \cite{rud:1}, pp 97-130.) In this paper, we shall provide a generalisation of this result and show in fact that one can obtain sets via complex oscillations which are linearly independent over the field of recursive real numbers.  Moreover, all the elements in these images are non-computable. We discuss the definability of these sets within the recursion-theoretic hierarchy, by exploiting the recursive isomorphism constructed in \cite{fo:2} between the Kolmogorov-Chaitin-Martin-L\" of random reals and the class of suitably encoded versions of  complex oscillations. \wedge
%\alpha \in KC.\]
%\label{th:arithmetic2}
%\end{theorem}
This is a $\Sigma_3^0$-representation, in effective descriptive set theory, of  countably random dense sets, independent and generic as explained above and given by the minimizers of Brownian motions which are encoded by $KC$-strings. \\
{\bf Remark}.  By specialising to a $\Delta_2^0$-element $\Omega_0$ in $KC$ (a Chaitin real), we thus find a $\Sigma_4^0$-predicate describing the local minimizers of the complex oscillation $\Phi(\Omega_0)$.
 
 The proofs of these theorems appear in Section \ref{section:local minimizers} of this paper.

%%%%%%%%%%%%%%%%%%%%%%%%%%%%%%%%%%%%%%%%%%%%%%%%%%%%%%%%%%%%%%%%%%%%%%%%%%%
 
\section{Effective descriptions of Brownian motion}

It is a daunting task to reflect sample path properties of Brownian motion (nowhere differentiability, law of the iterated algorithm, fractal geometry) into complex oscillations by defining these phenomena  in terms of the basic events in $B$ (see (\ref{eq:AP})) which is an effective basis for the uniform norm  topology in $C[0,1]$.  For this reason the author introduced in \cite {fo:1} another characterisation of the class $\mathcal{C}$ by using basic descriptions relative to effective Boolean subalgebras of the Borel algebra on $C[0,1]$. 

In order to describe this characterisation,  we follow \cite{fo:1} to define  an analogue of a $\Pi_2^0$ subset of $C[0,1]$ which is of constructive measure $0$. If $F$ is a subset of $C[0,1]$, we denote by $\overline{F}$ its topological closure  in $C[0,1]$ with the uniform norm topology. For $\epsilon > 0$, we let $O_\epsilon(F)$ be the $\epsilon$-ball \(\{f \in C[0,1]:\exists_{g \in F}\|f-g\| < \epsilon\}\) of $f$. (Here $\|.\|$ denotes the supremum norm.) 
We write $F^0$ for the complement of $F$ and $F^1$ for $F$.
\begin{definition}

A sequence $\mathcal{F}_0=(F_i:i < \omega)$ in $\Sigma$ is an \emph{effective generating sequence} if
\begin{enumerate}

\item for $F \in \mathcal{F}_0$, for $\epsilon > 0$ and $\delta \in \{0,1\}$, we have, for $G=O_\epsilon(F^\delta)$ or for $G=F^\delta$, that $W(\overline{G})=W(G)$,
\item there is an effective procedure that yields, for each sequence \( 0 \leq i_1 < \ldots < i_n < \omega\) and $k < \omega$ a binary rational number $\beta_k$ such that 
\[|W(F_{i_1} \cap \ldots \cap F_{i_n}) - \beta_k| < 2^{-k},\]
\item for $n,i < \omega$, a strictly positive rational number $\epsilon$ and for $x \in C_n$, both the relations $x \in O_\epsilon(F_i)$ and $x \in O_\epsilon(F_i^0)$ are recursive in $x,\epsilon,i$ and $n$, relative to an effective representation of the rationals.

\end{enumerate}
\label{def:effective}

\end{definition}
{\bf Remark}. This definition was motivated by the desire to have a class of basic statistical  events, which, firstly, are relevant to the practice of Brownian motion, and, secondly,  is such that  one can prove an effective version of the Donsker invariance principle and therefore, thirdly, to capture the entire class of complex oscillations.

If $\mathcal{F}_0 = (F_i:i < \omega)$ is an effective generating sequence and $\mathcal{F}$ is the Boolean algebra generated by $\mathcal{F}_0$, then there is an enumeration $(T_i:i < \omega)$ of the elements of $\mathcal{F}$ (with possible repetition) in such a way, for a given $i$, one can effectively describe $T_i$ as a finite union of sets of the form
\[F = F_{i_1}^{\delta_1} \cap \ldots \cap F_{i_n}^{\delta_n}\]
where $0 \leq i_1 < \ldots < i_n$ and $\delta_i \in \{0,1\}$ for each $i \leq n$. We call any such sequence $(T_i:i < \omega)$ a {\it recursive enumeration} of $\mathcal{F}$. We say in this case that $\mathcal{F}$ is {\it effectively generated} by $\mathcal{F}_0$ and refer to $\mathcal{F}$ as an {\it effectively generated algebra} of sets.

Let $\left(T_{i} : i <\omega\right)$ be a recursive enumeration of the algebra $\mathcal{F}$ which is effectively generated by the sequence $\mathcal{F}_{0}=\left(F_{i}:i<\omega\right)$ in $\Sigma$. It is shown in \cite{fo:1} that there is an effective procedure that yields, for $i,k<\omega$, a binary rational $\beta_{k}$ such that

$$|W\left(T_{i}\right)-\beta_{k}|<2^{-k},$$
in other words, the function $i \mapsto W(T_i)$ is computable.

A sequence $(A_n)$ of sets in $\mathcal{F}$ is said to be $\mathcal{F}$-{\it semirecursive} if it is of the form $(T_{\phi(n)})$ for some total recursive function $\phi: \omega \rightarrow \omega$ and some effective enumeration $(T_i)$ of $\mathcal{F}$. (Note that the sequence $(A_n^c)$, where $A_n^c$ is the complement of $A_n$, is also an $\mathcal{F}$-semirecursive sequence.) In this case, we call the union $\cup_nA_n$ a $\Sigma_1^0(\mathcal{F})$ set. A set is a $\Pi_1^0(\mathcal{F})$-set if it is the complement of a $\Sigma_1^0(\mathcal{F})$-set. It is of the form $\cap_n A_n$ for some $\mathcal{F}$-semirecursive sequence $(A_n)$. A sequence $(B_n)$ in $\mathcal{F}$ is a {\it uniform} sequence of $\Sigma_1^0(\mathcal{F})$- sets if, for some total recursive function $\phi:\omega^2 \rightarrow \omega$ and some effective enumeration $(T_i)$ of $\mathcal{F}$, each $B_n$ is of the form 
\[ B_n = \bigcup_m T_{\phi(n,m)}.\]
In this case, we call the intersection $\cap_nB_n$ a $\Pi_2^0(\mathcal{F})$-set. If, moreover, the  Wiener-measure of $B_n$ converges {\it effectively} to $0$ as $n \rightarrow \infty$, we say that the set given by $\cap_nB_n$ is a $\Pi_2^0(\mathcal{F})$-set of constructive measure $0$. 

The proof of the following theorem appears in \cite{fo:1}.
\begin{theorem}
Let $\mathcal{F}$ be an effectively generated algebra of sets. If $x$ is a complex oscillation, then $x$ is in the complement of every $\Pi_2^0(\mathcal{F})$-set of constructive measure $0$.
\label{theorem:quoteone}
\end{theorem}
This means, that every complex oscillation is, in an obvious sense, $\mathcal{F}$-Martin-L\" of random. The converse is also true.
\begin{definition} 
An effectively generated algebra of sets $\mathcal{F}$   is \emph{universal} if the class $\mathcal{C}$ of complex oscillations is definable  by  a   single $\Sigma_2^0(\mathcal{F})$-set,   the complement of which is a set of constructive measure $0$.
In other words, $\mathcal{F}$ is universal iff a continuous function $x$  on the unit interval is a complex oscillation iff $x$ is $\mathcal{F}$-Martin-L\" of random.
\label{definition:universal}
\end{definition} 

We introduce  two classes  of effectively generated algebras $\mathcal{G}$ and {$\mathcal{M}$ which are very useful for reflecting properties of one-dimensional Brownian motion into complex oscillations. %As was shown in  \cite{fo:1} and \cite{fo:2}, they are both universal in the sense just stated. 

Let $\mathcal{G}_0$ be a family of sets in $\Sigma$ each having a description of the form:
\begin{equation}
 a_1X(t_1) + \cdots + a_nX(t_n) \leq L
\label{eq:gaussone}
\end{equation}
or of the form (\ref{eq:gaussone}) with $\leq$ replaced by $<$, where all the $a_j,t_j\;(0 \leq t_j \leq 1)$ are non-zero computable real numbers, $ L$ is a recursive real number and $X$ is one-dimensional Brownian motion. 

%If  $\epsilon > 0$ and $G \in \Sigma$ is described by (\ref{eq:gaussone}), we have that \( O_\epsilon(G) \) is described by the inequality
%\begin{equation}
% a_1X(t_1) + \cdots +a_nX(t_n) < L + \epsilon\sum_j |a_j| 
%\label{equation:newgaussone}
%\end{equation}
%while $O_\epsilon(G^0)$ is given by
%\begin{equation}
 %a_1X(t_1) + \cdots +a_nX(t_n) > L - \epsilon \sum_j|a_j|. 
%\label{equation:newgausstwo}
%\end{equation}
We require that it be possible to find an enumeration $(G_i:i < \omega)$ of $\mathcal{G}_0$ such that, for given $i$, if $G_i$ is gi ven by (\ref{eq:gaussone}), we can effectively compute the sign,  and for every $n$, a rational approximation to each of  $a_j,t_j$ with error at most $1/n$.

%This has the implication that there is an effective procedure, $\Pi$, such that, for given $i, \epsilon, m$ with $i,m < \omega$ and $\epsilon$ a positive rational, the validity of (\ref{equation:newgaussone}) and (\ref{equation:newgausstwo}) can be decided by $\Pi$ when $G_i$ is given by (\ref{eq:gaussone}) and when $X \in C_m$. 

As in \cite{fo:2} in can be shown that $\mathcal{G}_0=(G_i:i<\omega)$ is an effective generating sequence in the sense of Definition \ref{def:effective}. The argument on p 325 of \cite{fo:2} holds verbatim for this slight generalisation. The associated effectively generated  algebra of sets $\mathcal{G}$ will be referred to as a {\it gaussian algebra}. 

It is shown in \cite{fo:1} that if $\mathcal{G}_0 $ is defined by  events of the form  (\ref{eq:gaussone}) with $n=1$ and $a_1=1$,  then the associated $\mathcal {G}$ is in fact  {\it universal} in the sense of Definition \ref{definition:universal}.

For a closed subinterval $I$ of the unit interval and a real number $b$, we write $[M(I) \geq b]$ for the event $[\sup\{ X(t):t \in I\} \geq b]$ and $[m(I) \leq b]$ for the event $[\inf\{X(t): t \in I\} \leq b]$, where $X$ is one-dimensional Brownian motion on the unit interval. We let $\mathcal{M}_0$ be the set of the events of the form $[M(I) \leq b]$ or $[m(I) \leq b]$ where $b$ is an arbitrary rational number and where $I$ is a subinterval of the unit interval with rational endpoints. It follows from the arguments on pp 434 - 438 in \cite{fo:1} that the elements of $\mathcal{M}_0$  can be effectively enumerated rendering $\mathcal{M}_0$  an effective generating sequence. 
We denote by $\mathcal{M}$ the Boolean algebra generated by $\mathcal{M}_0$. It is shown in \cite{fo:1} that $\mathcal{M}$ too is in fact universal.

We shall also make frequent use of the following result from \cite{fo:1} which is an easy  consequence of Theorem \ref{theorem:quoteone}. It is the analogue, for continuous functions, of the well-known fact that Kurtz-random reals contain the class of Martin-L\" of random reals.
\begin{theorem}
If $B$ is a $\Sigma_1^0(\mathcal{F})$ set and $W(B) = 1 $, then $\mathcal{C}$, the set of complex oscillations, is contained in $B$.
\label{theorem:quotetwo}
\end{theorem}

\section{Local minimizers of Brownian motion}
\label{section:local minimizers}

In this section we shall prove Theorems \ref{theorem:algorithm} and \ref{theorem:description}. A crucial remark is that the the local minimizers in subintervals of the unit interval of complex oscillations are uniquely determined:

\begin{proposition}
 If $x \in \mathcal{C}$, and $I_1,I_2$ are closed disjoint subintervals of the unit interval having rational endpoints, then
\[\inf_{t \in I_1} \;x(t) \neq \inf_{t \in I_2} \;x(t). \]
\label{proposition:fundamental}
\end{proposition}
\vspace*{-6mm}

The proof is a constructive version of the argument  on p 20 of \cite{per:1}. We shall need the following
\begin{lemma}
Let $Z_1, Z_2$ be independent real-valued random variables on some probability space with $Z_2$ having a non-atomic distribution. Then, almost surely, $Z_1+Z_2 \neq 0$.

\label{lemma:technical}
\end{lemma}
{\bf Proof}. For $i=1,2$, write $\mu_i$ for the distribution measure of $Z_i$. Then $Z_1+Z_2$ has the convolution product $\mu_1*\mu_2$ as its distribution measure, which will be non-atomic when $\mu_2$ is. Indeed, for a Borel set $A$ of real numbers,
\[ (\mu_1*\mu_2)(A) = \int_\mathbf{R}\mu_2(A-t)d\mu_1(t),\] 
and for $A=\{0\}$, it is the case that $\mu_2(A-t)=0$ for all $t$ (the measure $\mu_2$ being non-atomic). The result follows since $\mu_1*\mu_2(\{0\})$ is the probability of the event $[Z_1+Z_2=0]$.
\newline
{\bf Proof of Proposition \ref{proposition:fundamental}}. It is well known (see p20 of \cite{per:1}) that, under the hypotheses on $I_1,I_2$,  almost surely, $m(I_1) \neq m(I_2)$. Indeed, let $I_1=[a_1,b_1]$ and $I_2=[a_2,b_2]$  denote the lower and higher interval, respectively. Then the event $[m(I_1)=m(I_2)]$ is the same as
\[X(a_2)-X(b_1) =(m(I_1)-X(b_1)) -(m(I_2)-X(a_2)).\]

Since successive increments of Brownian motion are statistically independent, the random variable given by the expression on the right-hand side of this equation is independent from the the random variable on the left-hand side while the latter is non-atomic, being absolutely continuous with respect to Lebesgue measure. It follows from the preceding lemma that $m(I_1)\neq m(I_2)$ almost surely.

 The event $[m(I_1)\neq m(I_2)]$ is described by the following  $\Sigma_1^0(\mathcal{M})$ event of Wiener measure one:
\[\exists_{r \in \mathbf{Q}}\; \big(m(I_1) <r <m(I_2)\big) \vee \big(m(I_2) <r <m(I_1)\big).\]
The proposition follows from Theorem \ref{theorem:quotetwo} with $\mathcal{F}=\mathcal{M}$.\par
The proposition has the following
\begin{corollary}
For every complex oscillation, for every dyadic interval $I$ in the unit interval, there is a unique point in $I$ where the minimum of $x$ is assumed.
\label{corollary:essential}
\end{corollary}
We shall refer to this point as the {\em minimizer} of $x$ in $I$. We shall also need
\begin{lemma}
 If $x \in \mathcal{C}$ and $d_1, d_2$ are distinct rational numbers in the unit interval, then $x(d_1) \neq x(d_2)$.
\label{lemma:2}
\end{lemma}
{\bf Proof}.
 Indeed, for a one-dimensional Brownian motion $X$, the random variable $X(d_1)-X(d_2)$ is normal  with variance $|d_1-d_2|$ and is therefore non-atomic being absolutely continuous with respect to Lebesgue measure. Consequently, almost surely, $X(d_1) \neq X(d_2)$.

Moreover, the almost sure event $[X(d_1) \neq X(d_2)]$ has a $\Sigma_1^0(\mathcal{G})$ description with respect to a suitable gaussian algebra $\mathcal{G}$.  The description is given by the predicate 
$$\exists_{ r \in \mathbb{Q}^+}\;|x(d_1)-x(d_2)|>r$$
over $C[0,1]$.
Next apply Theorem \ref{theorem:quotetwo} with $\mathcal{F}=\mathcal{G}$.\newline
{\bf Remark}. The preceding argument can very easily be adapted to show that each complex oscillation is {\it injective} when restricted to the computable reals in  
the unit interval.

A consequence of Proposition \ref{proposition:fundamental} is that  for $x \in \mathcal{C}$, one can associate, with every dyadic subinterval $I$ of the unit interval, the unique real number $\tau_I^x \in I$  which is the local minimizer of $x$ in the interval $I$. By using this fact  we shall now show that one can find a measurable mapping $\mathcal{N}: KC \rightarrow (0,1)^\infty_{\neq}$ which upon composition with the projection $\pi :(0,1)^\infty \rightarrow (0,1)^\infty_{\neq}/S_\infty$ yields the strongly random set $\mathcal{MIN}:KC \rightarrow (0,1)^\infty_{\neq}/S_\infty.$

To define $\mathcal{N}$ we firstly note that it follows from L\' evy's arcsine law that the distribution of the variables $\tau_I^x$ are all absolutely continuous with respect to Lebesgue measure. (See, for example \cite{per:1}.) 

In fact, it follows from L\' evy's arcsine law that, if $I=[a,b]$ and $a < \alpha < \beta <b$ then

$$W(\alpha < \tau^x_I < \beta) =\frac{1}{\pi}\int_{\frac{\alpha-a}{b-a}}^{\frac{\beta-a}{b-a}}\frac{dt}{\sqrt{t(1-t})}.$$

 In particular, for a recursive real number $r$ in and a dyadic subinterval $I$ of the unit interval, the event $A$ given by

$$ x \in A \Leftrightarrow \tau_I^x =r$$
is such that $W(A)=0$. Moreover, writing $\mathbb{D}$ for the set of dyadic rationals in the unit interval: 

$$x \in A \Leftrightarrow \forall_{q \in \mathbb{D}}\;\; x(r) \leq x(q),$$ 
which means that $A$ has a $\Pi_1^0(\mathcal{G})$ description relative to a suitable gaussian algebra $\mathcal{G}$. It again follows from Theorem \ref{theorem:quotetwo} that $A$  contains no complex oscillations. 

Therefore, writing $\mathbb{R}_r$ for the field of recursive real numbers, we have:

\begin{theorem}
If $\alpha \in KC$ then $$\verb=MIN=(\Phi(\alpha)) \cap \mathbb{R}_r = \emptyset.$$
\label{theorem:final}
\end{theorem}

We now define the mapping $\mathcal{N}: KC \rightarrow (0,1)^\infty_{\neq}$ in stages. At stage $0$ we select the unique minimizer of $\Phi(\alpha)$ in the unit interval. At stage $n$, we first select $$\big(\tau_{[\frac{k}{2^n},\frac{k+1}{2^n}]}^{\Phi(\alpha)}:0 \leq k < 2^n\big)$$ and only add the local minimizers that haven't been selected at an earlier stage. The enumeration is well-defined for the minimizers are all, by Theorem \ref{theorem:final}, not endpoints of the intervals.\\
{\bf Remark.} It would be interesting to know, whether, for $\alpha \in KC$ it is the case that $$\verb=MIN=(\Phi(\alpha)) \subset KC.$$
{\bf Proof of Theorem \ref{theorem:algorithm}}. In the sequel we shall, for $\alpha \in KC$, denote the function $\Phi(\alpha)$ also by $x_\alpha$.

 Let $I$ be a fixed dyadic interval.  For $n \geq 1$, set
\[ D_n =\{\frac{k}{2^n}: 0 \leq k < 2^n \wedge [\frac{k}{2^n},\frac{k+1}{2^n}] \subset I\}.\]
For $d \in D_n$, set $I_d=[\frac{k}{2^n},\frac{k+1}{2^n})$, when $d=\frac{k}{2^n}$. Note that, for each $\beta \in I$, there is some $N$ such that $\beta \in \cup_{d \in D_n}I_d$ for all $n \geq N$.

It follows from Lemma \ref{lemma:2} and  Theorem \ref{theorem:isomorphism} that for $\alpha \in KC$  and dyadic rationals $d_1,d_2$, the relation $x_\alpha(d_1) < x_\alpha(d_2)$ is decidable in $\alpha, d_1,$ and $d_2$. It is because we know that $x_\alpha(d_1) \neq x_\alpha(d_2).$ (Lemma \ref{lemma:2}.)

Fix $\alpha \in KC$ and write $x_\alpha$ for the associated complex oscillation. The sequence $T=(t_n)$ is computed  by the prescription that for all $n$ with $D_n$ nonempty, we let $t_n$ be the unique element of $D_n$ such that $x_\alpha(t_n) < x_\alpha(d)$ for all $d \in D_n$ with $ d \neq t_n$. In view of Theorem \ref{theorem:isomorphism},  the sequence $T$ is computable from $\alpha$.

Let $(t_{n_k})$ be any convergent subsequence of $T$ with limit $\nu$, say. For given $\eta >0$, we have for all $k$ sufficiently large ($\geq L$, say) that $\nu \in [t_{n_k}- \eta, t_{n_k} + \eta]$.  Fix $\beta\in I$. Next choose $L_1 \geq L$ such   for  $k\geq L_1$ we can find some $d_k$  in $D_{n_k}$ such that $\beta \in I_{d_k}$.  

For $k \geq L_1$
\[ x_\alpha(\nu) -x_\alpha(\beta) =x_\alpha(\nu)- x_\alpha(t_{n_k})+ x_\alpha(t_{n_k}) - x_\alpha(d_k) +x_\alpha (d_k) -x_\alpha(\beta),\]
and, since, by construction,
\[x_\alpha(t_{n_k}) \leq x_\alpha(d_k),\]
the difference  $x_\alpha(\nu) -x_\alpha(\beta)$ can be made to be arbitrarily small by first choosing $\eta$ sufficiently small and then $k$ sufficiently large. We conclude that  $x_\alpha(\nu) \leq x_\alpha(\beta)$. In particular, 
\[x_\alpha(\nu) = m(I).\]

 Recall that $x_\alpha$ has a unique minimizer in $I$.
Hence  all the convergent subsequences of $T$ have the same limit.  We can therefore conclude that $T$ is a convergent sequence converging to the unique point in $I$ where the minimum of $x_\alpha$ on $I$ is assumed. This concludes the proof of the theorem.\newline
{\bf Remark.} Even though the construction of the sequence $(t_i:\;i \geq 0)$ in the theorem is effective relative to $\alpha \in KC$, the proof renders no information on the rate of convergence to the local minimizer in the dyadic interval. This problem will be addressed in a sequel of this paper (in collaboration with George Davie where it will be shown how it can be uniformly computed from the incompressibility coefficient of $\alpha  \in KC$).

We again write $\mathbb{D}$ for the dyadic rationals in the unit interval.
For the proof of the Theorem \ref{theorem:description}, we shall need the following
\begin{proposition}
The relations $x_\alpha(\mu) < x_\alpha(t)$  and $x_\alpha(\mu) > x_\alpha(t)$ are each $\Sigma_2^0$ in $\alpha \in KC,\;\mu \in \{0,1\}^\mathbb{N}$ and $t \in \mathbb{D}$.
\label{proposition: define}
\end{proposition}

To prove this Proposition, we first discuss the following Lemma:
\begin{lemma}
There is a uniform algorithm that, having access to an oracle for $\alpha \in KC$, will \emph{decide} whether 
$$\Phi(\alpha)(t) < q,$$
for $t \in \mathbb{D}$ and $q \in \mathbb{R}_r$.
\label{lemma:decide}
\end{lemma}
{\bf Proof:} It follows from \cite{fo:1} that, under the above hypotheses on $\alpha, t$ and $q$ 
$$\Phi(\alpha)(t)  \neq q.$$

Since $$\Phi(\alpha)(t) > q \Leftrightarrow \exists_n \overline{\Phi(\alpha)(t)}(n) > q,$$ the inequality
$\Phi(\alpha)(t) > q$ can be algorithmically affirmed if true. (This a direct consequence of Theorem \ref{theorem:isomorphism}.)

To affirm the inequality $\Phi(\alpha)(t)< q $ we need only apply (\ref{equation:fundamental symmetry}) and note that  $$\Phi(\alpha)(t)< q \Leftrightarrow \Phi(\hat{\alpha})(t)> -q,$$
to conclude the proof of the lemma.
It is shown in \cite{fo:2} that every complex oscillation is everywhere  $\beta$-H\" older continuous for any $0 < \beta < \frac{1}{2}$. 
The proof of of the $\Sigma_2^0$-definability of the relation $x_\alpha(\mu)<x_\alpha(t)$ now follows from this observation together with Lemma \ref{lemma:decide} which allows one to infer that:
$$x_\alpha(\mu)<x_\alpha(t) \Leftrightarrow \exists_{q \in \mathbb{Q}}\exists_k\forall_{L \geq k}\; x_\alpha(\overline{\mu}(L)) < q + \frac{1}{2^{L/3}}\wedge q < x_\alpha(t).$$
The $\Sigma_2^0$-definability of $x_\alpha(\mu)> x_\alpha(t)$ now follows from symmetry. (Replace $\alpha$ by $\hat{\alpha}$.)
\newline
{\bf Proof of Theorem \ref{theorem:description}}. Using the notation in the proof of the preceding theorem, define, for $\mu$ in the Cantor space $\{0,1\}^\mathbb{N}$,  and for $\alpha \in KC$ , the predicate $I_d(\mu, \alpha)$ to mean that the real in the unit interval corresponding to $\mu$ is the unique minimum on $I_d$ of $x_\alpha$. Here again, we have set $I_d=[\frac{k}{2^n},\frac{k+1}{2^n})$, when $d=\frac{k}{2^n}$. 
Note that for $\alpha \in KC$.
\[I_d(\mu, \alpha) \Leftrightarrow \mu \in I_d \wedge \forall_{t \in I_d\cap\mathbb{D}}\;[x_\alpha(\mu) < x_\alpha(t)].\] Since every local minimizer of a complex oscillation is a non-dyadic number, we can replace $[x_\alpha(\mu) < x_\alpha(t)]$ by $[x_\alpha(\mu) \leq x_\alpha(t)]$ in the definition of $I_d(\mu,\alpha)$. It therefore follows from Proposition \ref{proposition: define} that 
 the predicate $I_d(\mu, \alpha)$ is a $\Pi_2^0$-formula in $\mu$ and $\alpha$. Writing again  $\mathcal{N}(\alpha)$ for the set of local minimizers of $x_\alpha$, we find,
$$ \mu \in \mathcal{N}(\alpha) \leftrightarrow \exists_{d \in \mathbb{D}}\;I_d(\mu, \alpha).$$
 This is a $\Sigma_3^0$-formula in $\mu$ and $\alpha$. Finally note that the set $KC$ is $\Sigma_2^0$-definable. \newline
{\bf Remark:} In view of Tsirelson's Theorem \cite{tsi:1} (see Theorem \ref{theorem:tsirelson} above), it is an interesting problem, to characterise, for $x \in \mathcal{C}$,  the Borel sets $B$ of Lebesgue measure $0$ that are disjoint from $\verb=MIN=(x)$. This is not always the case. For instance, if $B=\{z\}$ where $z$ is the minimum of $x$ on the unit interval then of course $B$ will intersect the local minimisers of $x$.  On the other hand, if $B=Z_x$, the zero set of $x$, then $B$ does have Lebesgue measure $0$ and will be disjoint from $\verb=MIN=(x)$. To see this, note that
if $X$ is a continuous version of one-dimensional Brownian motion, then, almost surely, no local minimum of $X$ will be a zero of $X$.  For otherwise, there will be a neighbourhood of some zero of $X$ containing no other zeroes of $X$, which contradicts the well-known fact that the zero set of $X$ is almost surely perfect (the zero set being, for instance, almost surely,  a set of non-zero Hausdorff dimension). It follows that, we have, for each interval $I$ in the unit interval, almost surely
\[\exists_{r \in \mathbf{Q^+}}\; m(I) < -r \vee m(I) >r.\]
This is, for each closed interval $I$ with rational endpoints,  a $\Sigma_1^0(\mathcal{M})$ event of full Lebesgue measure and is consequently reflected in every complex oscillation. We conclude that
if $x$ is a complex oscillation, then
\[Z_x \cap \verb=MIN=(x) = \emptyset.\]

%\end{document}
%%%%%%%%%%%%%%%%%%%%%%%%%%%%%%%%%%%%%%%%%%%%%%%%%%%%%%%%%%%%%%%%%%%%%%%%%%%%%%%

\end{document}